\documentclass{aa}  

\usepackage{graphicx}
\usepackage{txfonts}

\usepackage{hyperref}
\newcommand{\dd}{\mathrm{d}}

\newcommand{\hMsun}{h^{-1}\mathrm{M}_\odot}
\newcommand{\ichMpc}{h^{3}\mathrm{Mpc}^{-3}}
\newcommand{\hMpc}{h^{-1}\mathrm{Mpc}}
\newcommand{\ihMpc}{h\mathrm{Mpc}^{-1}}

\newcommand{\qq}{\vec{q}}
\newcommand{\xx}{\vec{x}}

\newcommand{\vv}{\vec{v}}
\newcommand{\zz}{\vec{z}}

\usepackage{placeins}
\usepackage{amsfonts}
\usepackage{color}
\usepackage[dvipsnames]{xcolor}

\begin{document}

   \title{Modelling of post-reconstruction galaxy clustering with a hybrid bias expansion}
\subtitle{}

\titlerunning{Post-reconstruction hybrid bias model} \authorrunning{V. Izzo Dominguez et al.}


   \author{Vicente Izzo Dominguez\inst{1}        \fnmsep\thanks{\email{vicente.izzo@usm.cl}}
    \and Raul E. Angulo\inst{2,3} \fnmsep\thanks{\email{reangulo@dipc.org}}
    \and Antonio D. Montero-Dorta\inst{4}
    \fnmsep\thanks{\email{antonio.montero@usm.cl}}
    \and Marcos Pellejero-Ib\'a\~nez\inst{5}
        }

 \institute{Departamento de F\'isica, Universidad T\'ecnica Federico Santa Mar\'ia, Av. España 1680, Valparaíso, Chile
        \and
       Donostia International Physics Center (DIPC), Paseo Manuel de Lardizabal, 4, 20018, Donostia-San Sebasti\'an,
Gipuzkoa, Spain.
        \and 
        IKERBASQUE, Basque Foundation for Science, E-48013, Bilbao, Spain.
        \and
        Departamento de F\'isica, Universidad T\'ecnica Federico Santa Mar\'ia, Avenida Vicu\~na Mackenna 3939, San Joaqu\'in, Santiago, Chile.
        \and
        Institute for Astronomy, University of Edinburgh, Royal Observatory, Blackford Hill, Edinburgh EH9 3HJ, UK}
   \date{Received September 30, 20XX}

 
  \abstract
    {Density-field reconstruction sharpens the baryon acoustic oscillation (BAO) feature and recovers part of the cosmological information that nonlinear evolution removes from 2-point statistics. Using the reconstructed clustering beyond BAO measurements requires a model that captures both nonlinear evolution and galaxy formation physics.}
    {We extend the hybrid Lagrangian bias expansion to post-reconstruction galaxy clustering in configuration space. We validate the model, determine the range of scales on which is reliable, and quantify the cosmological information gained by combining pre- and post-reconstruction statistics.}
    {We apply a single reconstruction displacement to each Eulerian basis field of a second-order hybrid expansion, so that the post-reconstruction model depends on the same bias parameters as the pre-reconstruction one. We fit the correlation-function multipoles of subhalo-abundance matching mocks in real and redshift space, for the RecIso and RecSym conventions, and use a Fisher analysis to forecast constraints on $\sigma_8$ and $\Omega_{\rm m}$.}
    {The model reproduces the pre- and post-reconstruction multipoles for $r>35\,h^{-1}\mathrm{Mpc}$ with a single set of Lagrangian bias parameters. For a $3\,h^{-3}\mathrm{Gpc}^3$ volume at $z\simeq0$ with $\bar n=5\times10^{-4}\,h^3\mathrm{Mpc}^{-3}$, post-reconstruction clustering (POST) alone tightens the marginalised constraint on $\Omega_{\rm m}$ by a factor of $\sim1.3$ relative to pre-reconstruction clustering (PRE), but slightly weakens that on $\sigma_8$ when the reconstruction bias is inferred from the observed clustering amplitude. Combining PRE and POST reduces both marginalised uncertainties by about one third and increases the figure of merit by a factor of $\sim2.2$, mainly by breaking degeneracies between bias and cosmology.}
    {Pre- and post-reconstruction clustering contain complementary cosmological information and can be modelled within the same hybrid Lagrangian bias expansion. The framework is well suited to emulation and future joint full-shape analyses of galaxy surveys.}


   \maketitle
\nolinenumbers

\section{Introduction}

The large-scale distribution of galaxies is one of the primary probes of modern cosmology: it encodes the origin and growth of structure, the expansion history of the Universe, and the nature of gravity \citep{Peebles1980,Dodelson2020}. In particular, baryon acoustic oscillations (BAOs) provide a robust standard ruler and have become a cornerstone of cosmological analyses \citep{Eisenstein2005, Cole2005, Alam2021, DESI2024}. 

Nonlinear evolution, however, couples initially independent Fourier modes and transfers part of the sensitivity to cosmological parameters from the power spectrum into higher-order correlators \citep{Bernardeau2002}. Together with redshift-space distortions (RSD), it also smear the BAO signal, reducing its constraining power
\citep{Eisenstein2007,Crocce2008,angulo2008detectability, Matsubara2008}. Higher-order correlations and field-level inference can, in principle, recover this lost constraining power \citep[e.g.][]{Slepian2017, GilMarin2017, PhilcoxIvanov2022, JascheWandelt2013, Schmidt2020EFT, Nguyen_2024}. However, these approaches are computationally demanding and require complex modelling, which so far has limited their adoption in survey analyses.

BAO reconstruction offers a simpler alternative \citep{Eisenstein2007}. The central idea is to estimate the large-scale displacement field and partially reverse it by moving galaxies back along the estimated path. This sharpens the BAO feature and recovers most of the lost constraining power \citep{Schmittfull_2015}. Depending on the reconstruction convention, BAO reconstruction also modifies the large-scale redshift-space anisotropy \citep{Padmanabhan2012, white2015, Seo2016}. Post-reconstruction BAO measurements are now standard and have significantly improved the precision of the BAO scale in SDSS \citep{Padmanabhan2012}, BOSS \citep{Alam2017}, eBOSS \citep{Alam2021}, and DESI \citep{DESI2024_BAO, DESI2025}. 

Full-shape analyses could exploit reconstruction in the same way. \citet{wang2024extracting} showed that jointly analysing the pre- and post-reconstruction power spectra and their cross-power spectrum, $P_{\rm pre}$, $P_{\rm post}$, and $P_{\rm cross}$, captures part of the constraining power otherwise encoded in higher-order statistics. Applied to DESI DR1 luminous red galaxies (LRGs), adding $P_{\rm post}$ and $P_{\rm cross}$ to $P_{\rm pre}$ tightened the $\sigma_8$ constraint by 18-27\% \citep{wang2026nonlinear}. In $w$CDM, combined with cosmic microwave background (CMB) data, it tightened the constraint on $w$ by 5-15\%. 

Two families of models have been developed to exploit post-reconstruction clustering. The first is perturbative, where models were first formulated in the Zel'dovich approximation \citep{Padmanabhan2009, Noh2009, white2015,Chen2019} and later extended to one-loop order and to biased tracers in redshift space \citep{Hikage2017}. More recently, \citet{sugiyama2024developing} developed an infrared-resummed model for $P_{\rm post}$ and $P_{\rm cross}$, and \citet{sugiyama2024discreteness} added reconstruction-specific discreteness effects and shot noise. \citet{zhang2025cosmological} further constructed one-loop redshift-space models for all three spectra of the matter field, and \citet{sugiyama2025unified} developed a Lagrangian formalism that consistently treats galaxy bias, RSD, and reconstruction. However, these models are restricted to the large scales where perturbation theory remains accurate, and the operational details of the reconstruction algorithm are hard to capture analytically. 

The second family of models are based on numerical simulation, which reach a much wider range of scales and can mimic the reconstruction algorithm directly. For instance, \citet{wang2024emulating} built an emulator for post-reconstruction spectra using the Dark Quest simulations and a halo occupation distribution (HOD) model. However, these predictions depend on the assumed galaxy-halo connection, and HODs rely on simplifying assumptions that are not fully supported by state-of-the-art galaxy formation simulations \citep{contreras2024validating, ChavesMontero2023, wang2026stresstesting}.

The hybrid Lagrangian bias model combines the strengths of perturbative and simulation approaches \citep{Modi2020, kokron2021, Zennaro2023, pellejero2023}. The model adopts a perturbative expansion for the galaxy--matter connection and advects each term with a nonlinear displacement field measured in $N$-body simulations. Therefore, it inherits nonlinear dynamics without assuming a model for the galaxy--halo connection, which is instead absorbed into a small number of Lagrangian bias coefficients. Relations among these coefficients have been calibrated across large sets of halo and galaxy samples, providing physically motivated priors \citep{Zennaro2022priors}.

The hybrid model implemented in the \textit{BACCO Project} \citep{Angulo2021} has been extensively validated. It has recovered unbiased cosmological constraints in the nonlinear regime in the Beyond-2pt challenge and in recent \textit{Euclid} comparisons \citep{Krause_2025, Camacho-Quevedo}, and has yielded competitive $\sigma_8$ constraints from the BOSS full-shape power spectrum \citep{PellejeroIbanez2024}. More recently, it has been validated for the bispectrum \citep{ibanez2026validation}. These studies have primarily focused on Fourier-space statistics; here, we extend the validation to configuration space and after reconstruction.

In the context of post-reconstruction, a hybrid model would offer four advantages: it can i) extend analyses to smaller scales than perturbation theory allows, ii) describe pre- and post- clustering within a single self-consistent framework, iii) remain agnostic to the details of galaxy formation, and iv) implement the operational choices of the reconstruction algorithm directly, because its predictions are three dimensional fields.  

In this paper, we develop and validate such a hybrid model for post-reconstruction clustering. We apply the reconstruction displacement directly to each Eulerian basis field of the expansion, so that the post-reconstruction correlation function is a bilinear combination of a compact set of reconstructed basis correlations, weighted by the same Lagrangian bias parameters that describe pre-reconstruction clustering. We test the model against subhalo abundance-matching mocks built from the BACCO simulations at $z\simeq0$, in real and redshift space, and for two standard reconstruction conventions, and forecast the constraints obtained by combining both statistics. Hereafter, we refer to the correlation functions measured before and after reconstruction as PRE and POST, and to fits that describe both with a single set of Lagrangian bias parameters as JOINT.

This paper is organised as follows. In Sect.~\ref{sec:model}, we present the hybrid Lagrangian bias model, its extension to the reconstructed fields, and the resulting two-point estimators. Section~\ref{sec:simulations} describes the simulations, galaxy catalogues, and emulated realisations, together with our measurements and fitting procedure. Section~\ref{sec:results_real_redshift} validates the model in real and redshift space. In Sect.~\ref{sec:fisher}, we forecast cosmological constraints PRE, POST, and their combination. Finally, we summarise our conclusions in Sect.~\ref{sec:conclusions}.

\section{Modelling the post-reconstruction statistics}
\label{sec:model}

We begin by reviewing the hybrid Lagrangian bias expansion (Sect.~\ref{subsec:hybrid}) and the BAO reconstruction algorithm (Sect.~\ref{subsec:reconstruction}). We then describe our model for post-reconstruction clustering (Sect.~\ref{subsec:postrecon_model}).

\subsection{The hybrid bias expansion}
\label{subsec:hybrid}

We adopt the hybrid Lagrangian bias framework developed in \citet{Zennaro2023} and \citet{pellejero2023}, in which a perturbative Lagrangian bias expansion is advected with the fully nonlinear displacement. Following the BACCO implementation, we define the five Lagrangian basis operators as
\begin{equation}
    \left\{
    \mathcal{O}_0,\ldots,
    \mathcal{O}_4
    \right\}
    =
    \left\{
    1,\,
    \delta_{\rm L},\,
    \delta_{\rm L}^2-\langle\delta_{\rm L}^2\rangle,\,
    s^2-\langle s^2\rangle,\,
    \nabla^2\delta_{\rm L}
    \right\},
\end{equation}

\noindent where $1$ is the operator describing unweighted matter, $\delta_{\rm L}(\qq)$ is the linear matter overdensity in Lagrangian space, smoothed with a Fourier-space top-hat filter at $k_{\rm d} = 0.75\,\ihMpc$, and $s^2=s_{ij}s^{ij}$ with $s_{ij}(\qq) = \left( \partial_i\partial_j/\nabla^2- \delta^K_{ij}/3 \right) \delta_{\rm L}(\qq)$.

After advection with a simulation displacement, $\boldsymbol{\Psi}$, each operator defines an Eulerian basis field,
\begin{equation}
    [\mathcal{O}_i](\xx)
    =
    \int \dd^3\qq\,
    \mathcal{O}_i(\qq)\,
    \delta_D\!\left[
        \xx-\qq-\boldsymbol{\Psi}(\qq)
    \right].
\end{equation}
The galaxy density field is a linear combination of these fields,
\begin{equation}
    1+\delta_g(\xx)
    =
    \sum_{i=0}^{4}
    b_i[\mathcal{O}_i](\xx),
\end{equation}

\noindent with bias coefficients $\{b_0, b_1, b_2, b_{s^2}, b_{\nabla^2\delta}\}$ and $b_0\equiv1$. 

\begin{figure*}[!t]
    \centering
    \includegraphics[width=0.9\linewidth]{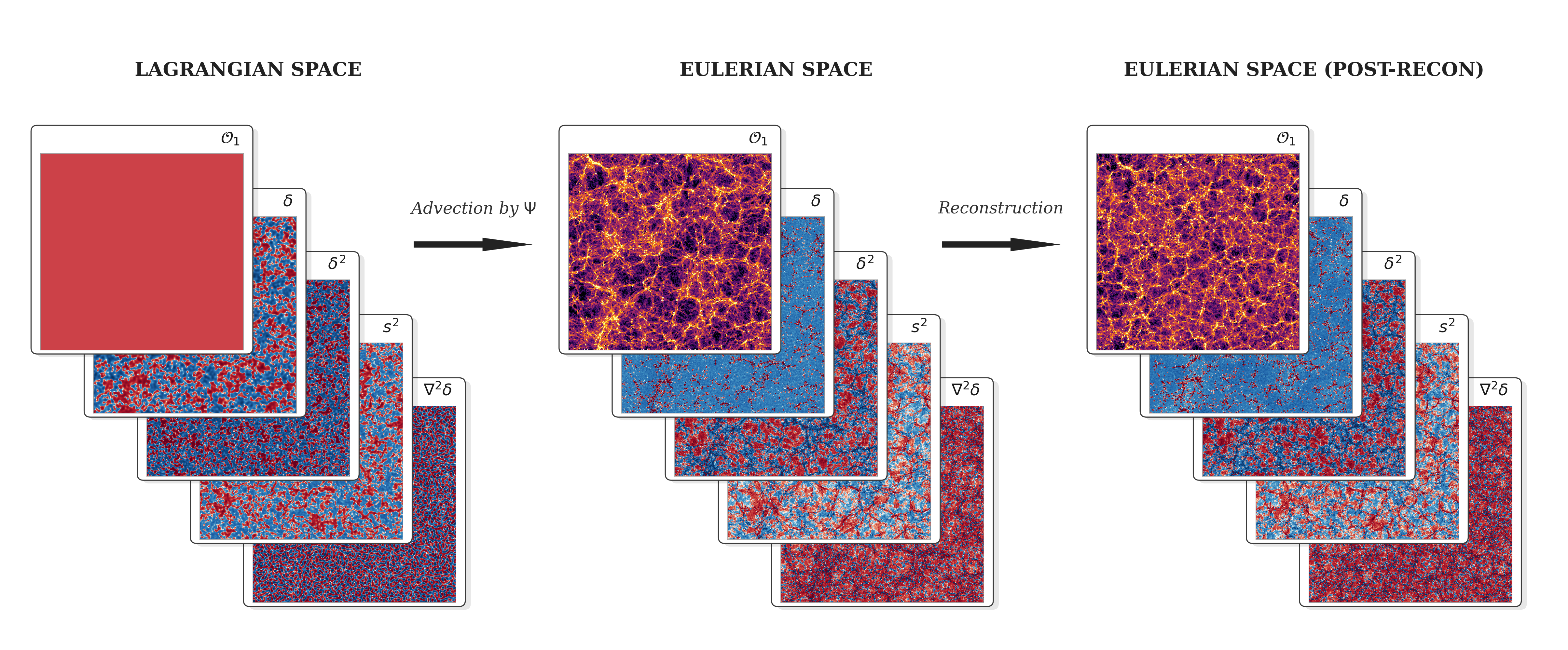}
    \caption{Workflow of the post-reconstruction hybrid model. The Lagrangian bias operators are advected to Eulerian space using the nonlinear displacement field $\boldsymbol{\Psi}$. A single reconstruction displacement $\boldsymbol{\Psi}^{(1)}$, estimated from the Eulerian matter density field with $b_{\rm rec}=1$, is then applied to the mock galaxy catalogue and to each Eulerian basis field. For a fixed reconstruction setting, $\boldsymbol{\Psi}^{(1)}$ is applied to all basis fields, so the reconstructed galaxy field remains a linear combination of the reconstructed operators with the same Lagrangian bias parameters as before reconstruction.}
    \label{fig:workflow}
\end{figure*}

We include redshift-space distortions at the field level by adding the line-of-sight velocity to the displacement,
\begin{equation}
    \boldsymbol{\Psi}^{\,s}(\qq)
    =
    \boldsymbol{\Psi}(\qq)
    +
    \frac{
        \vv(\qq)\cdot\hat{\zz}
    }{
        aH
    }\,
    \hat{\zz},
    \label{eq:rsd_displacement}
\end{equation}
where $\hat{\zz}$ denotes the line-of-sight direction in the plane-parallel approximation. Following \citet{pellejero2023}, particles inside haloes are assigned the centre-of-mass velocity of their host halo, and all other particles keep their own velocity. This captures large-scale coherent flows and the bulk motion of haloes. Here we do not include the relative motions of satellites within their hosts, which can be added through a Finger-of-God term. However, we expect their effect to be confined to separations below the fiducial minimum scale we use.

The galaxy two-point statistics are then bilinear combinations of the 5 auto- and 10 cross-correlations of the basis fields. In terms of $\xi_{ij,\ell}$, the $\ell$-th multipole of the cross-correlation between $[\mathcal{O_i}]$ and $[\mathcal{O_j}]$, we obtain: 

\begin{equation}
    \xi_{gg,\ell}(r)
    =
    \sum_{i\le j}
    (2-\delta^K_{ij})\,
    b_i b_j\,
    \xi_{ij,\ell}(r),
    \label{eq:xi_prerecon}
\end{equation}

\noindent We use the monopole ($\ell=0$), quadrupole ($\ell=2$), and hexadecapole ($\ell=4$).

A key feature of the model is that it predicts the galaxy overdensity as a three-dimensional field. Consequently, field-level operations such as reconstruction can be applied directly to the basis fields. The next section employs this property to develop our post-reconstruction model.

\subsection{Reconstruction algorithm}
\label{subsec:reconstruction}

We follow the standard reconstruction approach \citep{Eisenstein2007, Padmanabhan2012}, which consists of two steps. First, a displacement field is estimated from the smoothed galaxy density field. Second, galaxies (referred to as tracers) and a reference catalogue are moved along this displacement, defining a reconstructed density field.

\subsubsection{Estimating the displacement}
\label{sec:rec_displacement}

At linear order, the displacement that carries matter from its Lagrangian to its Eulerian position is the Zel'dovich displacement, $\nabla\cdot\boldsymbol{\Psi}^{(1)} = -\delta_{\rm m}$. Reconstruction estimates this field from the smoothed observed galaxy density field in redshift space, $\delta^{s}_{\rm in}$, and partially undoes it. 

The displacement $\boldsymbol{\Psi}^{(1)}$ is therefore obtained by solving the linearised continuity equation

\begin{equation}
  \nabla\cdot\boldsymbol{\Psi}^{(1)}
  + \frac{f_{\rm rec}}{b_{\rm rec}}\,
    \nabla\cdot\left[\left(\boldsymbol{\Psi}^{(1)}\cdot\hat{\boldsymbol{z}}\right)\hat{\boldsymbol{z}}\right]
  = -\frac{\delta^{s}_{\rm in}}{b_{\rm rec}},
  \label{eq:continuity}
\end{equation}

\noindent where $b_{\rm rec}$ and $f_{\rm rec}$ are the linear bias and growth rate assumed by the reconstruction. In real space, the second term is absent.

Our settings closely follow those adopted by the DESI Collaboration \citep{paillas2025}:

\begin{enumerate}[i)]

  \item We smooth the input field with a Gaussian kernel, $S(k)=\exp\!\left(-k^2\Sigma_{\rm sm}^2/2\right)$, with $\Sigma_{\rm sm}=15\,h^{-1}\mathrm{Mpc}$.
  
  \item We fixed $f_{\rm rec}=0.5103$, the linear growth rate of the cosmology and redshift of our simulation.
  
  \item We solve Eq.~\eqref{eq:continuity} using the iterative FFT method of \citet{burden2015}, as implemented in the \texttt{pyrecon} package%
  \footnote{\url{https://github.com/cosmodesi/pyrecon}}, with $n_{\rm iter}=4$ iterations on a mesh of $1080^3$ cells. 
\end{enumerate}

This choice is motivated by the convergence tests of \citet{chen2026reconstruction}, who found that the post-reconstruction power-spectrum multipoles are nearly insensitive to increasing the number of iFFT iterations beyond three. Moreover, our reconstruction has $f_{\rm rec}/b_{\rm rec}\simeq 0.51<1$, satisfying the convergence condition of the iFFT algorithm in the plane-parallel approximation.

\subsubsection{The reconstructed density field}
\label{sec:rec_field}

The next step is to move each tracer by: 

\begin{equation}
  \boldsymbol{d}_D = -\boldsymbol{\Psi}^{(1)}
  - f_{\rm rec}\left(\boldsymbol{\Psi}^{(1)}\cdot\hat{\boldsymbol{z}}\right)\hat{\boldsymbol{z}},
  \label{eq:dD}
\end{equation}

\noindent where the second term is absent in real space. This undoes the large-scale bulk flows that smear the BAO feature, but it also removes most of the large-scale clustering: on scales above $\Sigma_{\rm sm}$, the displaced tracer field $\delta_D$ retains only the Lagrangian-bias part of the large-scale clustering, $(b_1^E - 1)\,\delta_{\rm m}$, which is small for our sample. Shifting a uniform reference catalogue by $\boldsymbol{d}_D$ gives a field $\delta_S$ that contains only the density field generate by the displacement. Subtracting it restores the large-scale signal while keeping the sharpened BAO feature. Therefore, the reconstructed field is defined as:

\begin{equation}
  \delta_{\rm rec} \equiv \delta_D - \delta_S .
  \label{eq:delta_rec}
\end{equation}

Two reconstruction conventions exist, which differ in $\boldsymbol{d}_D$ \citep{white2015, Chen2019}:

\begin{itemize}

  \item \textbf{RecSym}: uses $\boldsymbol{d}_S=\boldsymbol{d}_D$ and preserves the large-scale redshift-space anisotropy.
  
  \item \textbf{RecIso}: uses $\boldsymbol{d}_S=-\boldsymbol{\Psi}^{(1)}$ and removes the large-scale redshift-space anisotropy at linear order.
  
\end{itemize}

In real space, the two conventions coincide, so we distinguish them only in redshift space.

For the fiducial mock and the hybrid model basis, we employ a reference catalogue given by dense quasi-random catalogue based on a Halton sequence \citep{Kerscher2022}. For the emulated realisations (Sect.~\ref{subsec:emulated}), we use a uniform lattice of $736^3$ points for the covariance suite and of $368^3$ points for the Fisher derivatives. A lattice avoids adding a stochastic contribution from the reference catalogue.

\subsection{The post-reconstruction model}
\label{subsec:postrecon_model}

The central idea of our approach is shown schematically in Fig.~\ref{fig:workflow}. A single displacement field $\boldsymbol{\Psi}^{(1)}$ is applied to the mock galaxy catalogue and to each of the five basis fields $\{[\mathcal{O}_i](\xx)\}$.  The reference catalogue is shifted by $\boldsymbol{d}_S$, according to the RecSym or RecIso convention (Sect.~\ref{sec:rec_field}).

For the reconstructed basis to be independent of the galaxy bias, we estimate $\boldsymbol{\Psi}^{(1)}$ from the matter field rather than from the galaxy field,
\begin{equation}
  \delta^{s}_{\rm in} = \delta^{s}_{m}, \qquad b_{\rm rec} = 1,
  \label{eq:matter_rec}
\end{equation}

\noindent where $\delta^{s}_{m}$ is the overdensity of the redshift-space number-operator field $[1]^s$. At linear order, this gives the same displacement as reconstructing the tracer field $\delta^s_g \simeq (b_1^E + f\mu^2)\,\delta_{\rm m}$ with $b_{\rm rec} = b_1^E$, provided $f_{\rm rec}$ equals the true growth rate. However, it neglects the nonlinear bias and shot-noise contributions of a real tracer field. We test this idealization in Sect.~\ref{subsec:robustness} by reconstructing the mock form its own galaxy field.

For a given displacement $\boldsymbol{d}$, reconstruction maps a density field to

\begin{equation}
  \mathcal{R}_{\boldsymbol{d}}[\rho](\boldsymbol{y}) =
  \int \mathrm{d}^3x\ \rho(\boldsymbol{x})\,
  \delta_D\!\left[\boldsymbol{y} - \boldsymbol{x} - \boldsymbol{d}(\boldsymbol{x})\right].
  \label{eq:R_operator}
\end{equation}

Each element moves by an amount independent of its weight, thus $\mathcal{R}_{\boldsymbol{d}}$ is linear in $\rho$. Defining $[\mathcal{O}_i]_r \equiv \mathcal{R}_{\boldsymbol{d}_D}\big[[\mathcal{O}_i]\big]$ gives

\begin{equation}
  \mathcal{R}_{\boldsymbol{d}_D}\!\left[\sum_{i=0}^{4} b_i\,[\mathcal{O}_i]\right]
  = \sum_{i=0}^{4} b_i\,[\mathcal{O}_i]_r .
  \label{eq:rec_linearity}
\end{equation}

Expanding the two-point function of $\delta_{\rm rec}$ (Eq.~\ref{eq:delta_rec}) with Eq.~\eqref{eq:rec_linearity} gives, for the even multipoles,

\begin{equation}
  \xi_{\rm rec} = \sum_{i\le j}(2-\delta^K_{ij})\,b_ib_j\,\xi^r_{ij}
  - 2\sum_i b_i\,\xi^r_{iS} + \xi^r_{SS},
  \label{eq:xi_rec_expanded}
\end{equation}

\noindent where $\xi^r_{ij}$, $\xi^r_{iS}$, and $\xi^r_{SS}$ are the auto- and cross-correlations of the mean-subtracted fields $[\mathcal{O}_i]_r$ and $\delta_S$. Since $b_0 \equiv 1$, the reference terms can be absorbed into the correlations containing the number operator:

\begin{equation}
  \tilde{\xi}^r_{ij} =
  \begin{cases}
    \xi^r_{00} - 2\,\xi^r_{0S} + \xi^r_{SS}, & i=j=0,\\
    \xi^r_{0j} - \xi^r_{jS}, & i=0,\ j\ge1,\\
    \xi^r_{ij}, & i,j\ge1,
  \end{cases}
  \label{eq:ref_corrected_basis}
\end{equation}

\noindent so that
\begin{equation}
  \xi^{\rm rec}_{gg,\ell}(r) = \sum_{i\le j}(2-\delta^K_{ij})\,b_ib_j\,\tilde{\xi}^r_{ij,\ell}(r).
  \label{eq:xi_rec_model}
\end{equation}

Equation~\eqref{eq:xi_rec_model} is the central result of this section: post-reconstruction clustering is a bilinear combination of 15 basis correlations, weighted by the same Lagrangian bias parameters as before reconstruction. The basis depends on cosmology and on the reconstruction settings ($\Sigma_{\rm sm}$, $f_{\rm rec}$, $b_{\rm rec}$), while galaxy formation enters only through the bias parameters. In practice, we measure the 21 auto- and cross-spectra among $\{[\mathcal{O}_i]_r, \delta_S\}$ on a common mesh and combine them according to Eq.~\eqref{eq:ref_corrected_basis}. 

\section{Numerical simulations and catalogues}
\label{sec:simulations}

\begin{figure*}[!t]
    \centering
    \includegraphics[width=0.8\linewidth]{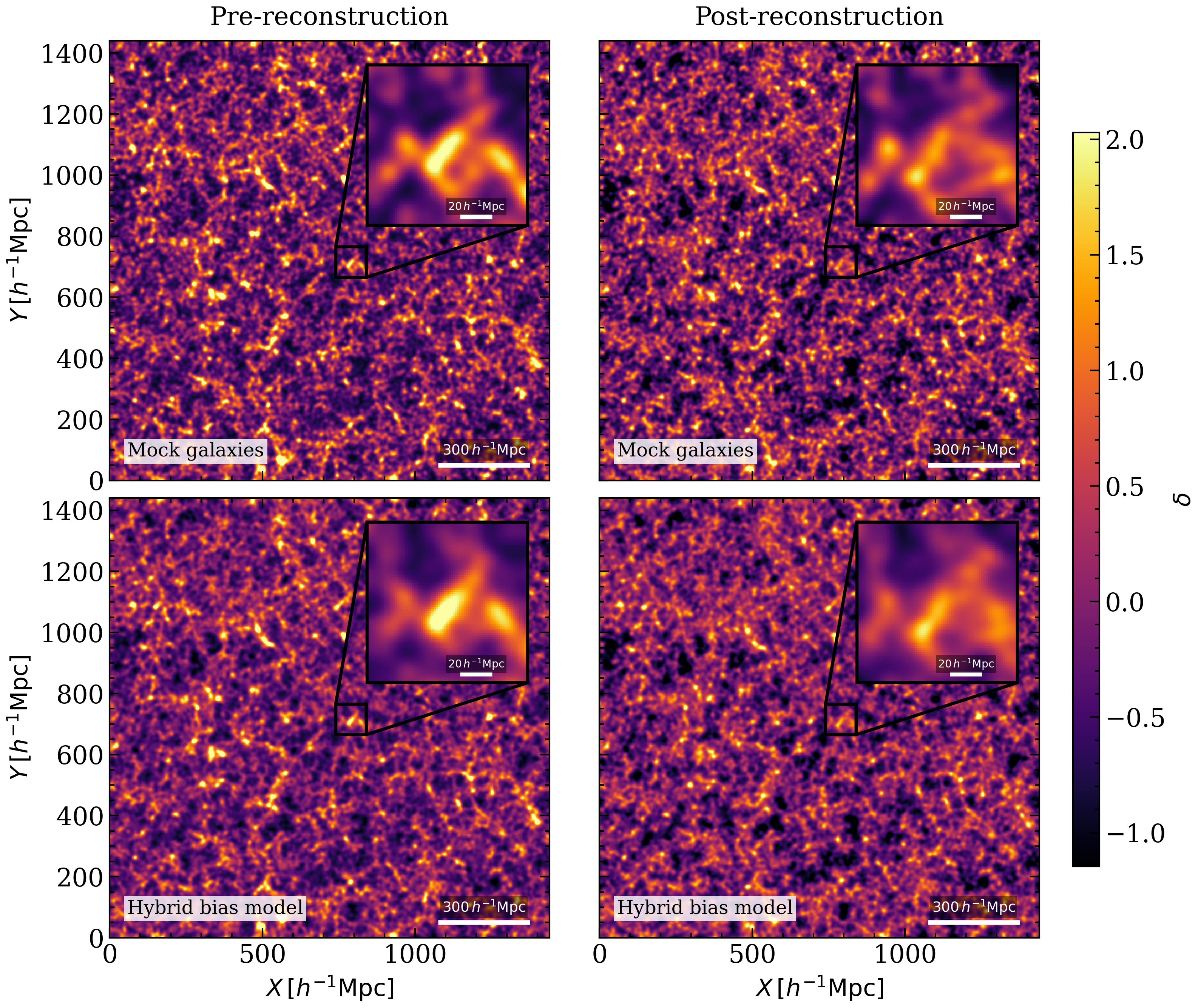}
    \caption{Projected redshift-space tracer density fields at $z\simeq0$ before (left) and after (right) RecSym reconstruction. The top row shows the mock galaxy catalogue and the bottom row the hybrid model, evaluated with the same best-fitting JOINT parameters. The post-reconstruction panels show $\delta_{\rm rec}=\delta_D-\delta_S$. The fields are projected along the line of sight over an $80\,\hMpc$ slab and smoothed with a Gaussian kernel of width $5\,\hMpc$. The colour scale is the same in every panel and indicates the overdensity. The insets enlarge a $100\times100\,h^{-2}{\rm Mpc}^2$ region. The mock catalogue has $\bar n=5\times10^{-4}\,\ichMpc$.}
    \label{fig:density_field}
\end{figure*}

\subsection{The BACCO simulations}
\label{subsec:bacco}

Our analysis is based on the \texttt{TheOne} simulation of the \textit{BACCO Project} \citep{Angulo2021}. BACCO consists of high-resolution gravity-only simulations that, combined with cosmology rescaling techniques \citep{AnguloWhite2010}, predict the nonlinear distribution of matter and biased tracers over a broad range of cosmologies. 

\texttt{TheOne} consists of a fixed-and-paired pair of simulations \citep{angulopontzen2016}, each following $4320^3$ particles in a periodic box of side $L=1440\,\hMpc$, which corresponds to a particle mass of $m_{\rm p}=3.2\times10^9\,\hMsun$. They were evolved with an updated version of \textsc{L-Gadget3} \citep{springel2005,Angulo2021}. We use the snapshot at $a = 1.0153$. Initial conditions were generated at $z_{\rm ini}=49$ using second-order Lagrangian perturbation theory. Averaging the two realisations strongly suppresses finite-volume fluctuations in the large-scale clustering \citep{chuang2019,maion2022}. This provides a low-noise fiducial data vector for validating our model. 

The simulation cosmology is
$\Omega_{\rm m}=0.307$,
$\Omega_{\Lambda}=0.693$,
$\Omega_{\rm b}=0.048$,
$h=0.677$,
$n_{\rm s}=0.9611$, and
$\sigma_8=0.9$. This is close to the \textit{Planck} 2018 best fit  \citep{planck2018}, except $\sigma_8$, which is higher to allow the simulation to be rescaled to a wider range of cosmologies. 

\subsection{Mock galaxy catalogues}
\label{subsec:mocks}

To validate our model, we construct mock galaxy catalogues using the extended subhalo abundance-matching model \texttt{SHAMe} \citep{contreras2021flexible}. \texttt{SHAMe} includes prescriptions for satellite tidal stripping and disruption, environmental dependence, and a parametrised relation between subhalo properties and star-formation rates. \citet{contreras2021flexible} showed that \texttt{SHAMe} reproduces the real- and redshift-space clustering of stellar-mass-selected galaxies in the TNG300 hydrodynamical simulation \citep{marinacci2018, naiman2018, nelson2018, pillepich2018, springel2018}.

We select galaxies by a property mimicking stellar mass, to give a number density $\bar n = 5\times10^{-4}\,\ichMpc$. This is comparable to that of spectroscopic surveys such as DESI luminous red galaxies, although at lower redshift. We analyse these catalogues in real and in redshift space. We measure the clustering statistics in each of the two \texttt{TheOne} realisations and average the results for the model validation and parameter fitting, while statistical uncertainties come from an ensemble of realisations we describe next. 

\subsection{Emulated mock suite}
\label{subsec:emulated}
We build a suite of emulated realisations with independent initial conditions, in boxes of the same size as TheOne, to estimate covariance matrices and the finite-difference derivatives required for the Fisher analysis (cf. Sect.~\ref{sec:fisher}). For each cosmology and initial Gaussian random field, we generate a Zel'dovich displacement field. We then apply the \texttt{map2map} emulator \citep{jamieson2022, pellejero2024map2map} at its native resolution, $N_{\rm mesh}=736$, to predict the corresponding nonlinear displacement and velocity fields. The bias operators are built and advected to Eulerian space, following the same procedure as for \texttt{TheOne}. 

\subsubsection{Covariances}
We generate $N_{\rm real}=600$ realisations at the fiducial cosmology. In each, we create a hybrid tracer field with the Lagrangian bias parameters $\left( b_1,\,  b_2,\, b_{s^2},\, b_{\nabla^2\delta} \right) = \left( 0.188,\, -0.319,\, 0.774,\, 0.675 \right) $. These values come from a preliminary fit to the fiducial sample and agree with the final PRE fits at the $\lesssim1.5\sigma$ level. 

To mimic a finite tracer number density, we first assign each Lagrangian particle a weight $ w(\qq)=\sum_i b_i^{\rm cov}\, \mathcal{O}_i(\qq)$, where $b_i^{\rm cov}$ are the coefficients above, and set negative weights to zero.\footnote{Approximately 20\% of particles have negative weights, with very little variation between realisations. Setting these weights to zero and renormalising $\lambda_p$ preserves the target mean number density but modifies the clustering amplitude of the discrete  tracer. The covariance is therefore designed to reproduce the simulation volume, the number density, the reconstruction procedure, and approximately the large-scale clustering of the sample.} We then draw the number of tracers associated with each particle from a Poisson distribution, $N_p\sim{\rm Poisson}(\lambda_p)$, with 

\begin{equation}
    \lambda_p =
    \max[w(\qq_p),0]\,
    \frac{\bar n V}
    {\displaystyle\sum_q \max[w(\qq_q),0]},
    \label{eq:poisson_sampling}
\end{equation}
\noindent where $\bar n=5\times10^{-4}\,\ichMpc$, matching our mock catalogue, and $V$ is the box volume. The resulting covariance therefore includes both finite-volume fluctuations and the stochastic contribution from tracer discreteness. 

PRE and POST are measured from the same realisations, so their auto- and cross-covariance can be estimated jointly. In each realisation, we i) draw a single Poisson realisation of the hybrid tracer catalogue and measure PRE from it; ii) estimate the reconstruction displacement from the corresponding matter field; and iii) apply that displacement to the same Poisson-sampled catalogue and measure POST. Thus, PRE and POST share the same tracer-discreteness noise. The reference catalogue is a uniform lattice, shifted according to the RecIso or RecSym convention. 

\subsubsection{Fisher derivatives}

For the finite-difference derivatives with respect to $\sigma_8$ and $\Omega_{\rm m}$, we run one fixed-and-paired pair of \texttt{map2map} realisations at each cosmological endpoint. We then averaging each pair to suppress sample variance and reduce noise in the numerical derivatives. We evaluate positive and negative variations at both the full and half step sizes, keeping the initial phases matched across cosmologies.


\subsection{Measurements and model fitting}
\label{sec:postmodel}
\begin{figure*}[!t]
    \centering
    \includegraphics[width=0.8\textwidth]{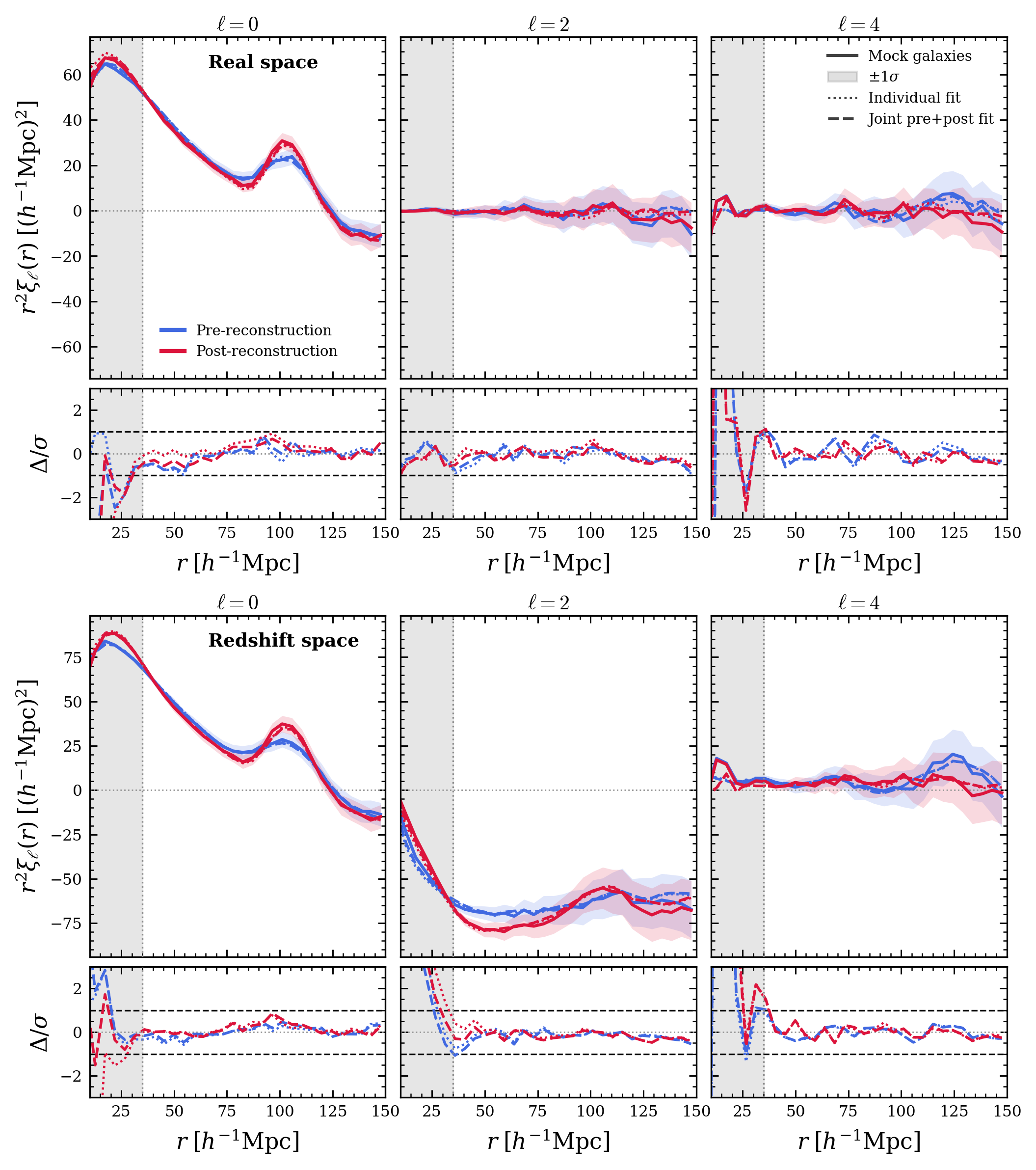}
    \caption{Correlation-function multipoles before (blue) and after (red) reconstruction, in real space (top) and in redshift space (bottom) for RecSym. Columns correspond, from left to right, to the monopole ($\ell=0$), quadrupole ($\ell=2$), and hexadecapole ($\ell=4$). Solid lines show the fixed-and-paired mock measurements; dotted lines show the best fits from the independent PRE and POST fits. Dashed lines show the JOINT fits, with a single set of bias parameters shared between PRE and POST. Shaded regions indicate the $1\sigma$ uncertainties estimated from the covariance ensemble. Lower panels show the residuals $(\mathrm{Mock}-\mathrm{Model})/\sigma$. The grey region marks $r < r_{\min}=35\,\hMpc$, which is excluded from the fits.}
    \label{fig:multipoles}
\end{figure*}

For the $N$-body mock and the hybrid-model predictions, we deposit the fields on a $N_{\rm mesh}=1080$ mesh using a cloud-in-cell mass assignment. We compute the auto- and cross-correlation functions by inverse Fourier transforming the deconvolved three-dimensional auto- and cross-power spectra on the mesh, and set the $k=0$ mode to zero. We then project $\xi(\mathbf{r})$ onto Legendre polynomials in 32 linear radial bins over $1<r<150\,\hMpc$.

Our data vector is $\mathbf{d} = \left\{\xi_0(r),\, \xi_2(r),\, \xi_4(r)\right\}$ restricted to $r > r_{\min}$, where our fiducial choice is $r_{\min} = 35\,\hMpc$ (justified in Sect.~\ref{sec:scalecuts}). The data vector has $N_d=75$ elements for PRE or POST alone and $N_d=150$ for JOINT. We determine the Lagrangian bias parameters $\boldsymbol{b}$ by minimising:

\begin{equation}
    \chi^2(\boldsymbol{b})
    =
    \left[
        \mathbf{d}-\mathbf{m}(\boldsymbol{b})
    \right]^{\rm T}
    C_{\rm H}^{-1}
    \left[
        \mathbf{d}-\mathbf{m}(\boldsymbol{b})
    \right],
    \label{eq:bias_chi2}
\end{equation}

\noindent where $\mathbf{m}(\boldsymbol{b})$ is the model described in Sect.~\ref{subsec:postrecon_model}. $C_{\rm H}^{-1}$ is the inverse covariance corrected for the finite number of realisations using the Hartlap factor \citep{hartlap2007},
\begin{equation}
    C_{\rm H}^{-1} = \frac{N_{\rm real}-N_{\rm d}-2} {N_{\rm real}-1}  \hat{C}^{-1},
    \label{eq:hartlap_fit}
\end{equation}

\noindent where $N_{\rm d}$ is the length of the corresponding data vector. With $N_{\rm real}=600$, the Hartlap factor is $0.87$ for PRE and POST, and $0.75$ for JOINT.

Since the model is quadratic in $\boldsymbol{b}$, the Jacobian $\mathbf{J}_b$ is analytic. The minimisation is performed using a multi-start trust-region nonlinear least-squares algorithm. We verified that independent optimisation methods recover the same minima. The parameter uncertainties in Sect.~\ref{sec:results_real_redshift} are the square roots of the diagonal of $(\mathbf{J}_b^{\rm T}\mathbf{C}_{\rm H}^{-1} \mathbf{J}_b)^{-1}$, at the best-fits, with  $\mathbf{J}_b$ defined as in Eq.~\ref{eq:jb}.

\section{Results in real and redshift space}
\label{sec:results_real_redshift}

We validate the model in four steps. We first compare the model and the mock at the level of the density field and of the correlation-function multipoles (Sect.~\ref{sec:accuracy}). We then determine the range of scales over which the fitted bias parameters are stable (Sect.~\ref{sec:scalecuts}), and test whether a single set of bias parameters describes pre- and post-reconstruction clustering simultaneously (Sect.~\ref{sec:singlebias}). Finally, we test the sensitivity of these results to the matter-driven reconstruction (Sect.~\ref{subsec:robustness}).

\subsection{Model accuracy}
\label{sec:accuracy}

Figure~\ref{fig:density_field} compares the mock galaxy field with the hybrid model in redshift space, before and after RecSym reconstruction, with the model evaluated at the best-fitting JOINT parameters. We see that reconstruction redistributes the density field in the same way for the mock and for the model, as expected from Eq.~\eqref{eq:rec_linearity}. The mock and the model agree remarkably well on large scales, which is expected as they share initial conditions. However, their agreement on small scales is not trivial: the model also reproduces the filamentary network and the compact overdensities of the mock, including within the zoomed region. 

Figure~\ref{fig:multipoles} compares the measured and predicted multipoles before and after reconstruction, in real space and in redshift space for RecSym (the RecIso multipoles are fitted equally well). The shaded bands show the statistical uncertainty of a single simulation volume, estimated from the covariance ensemble. Since the mock and the model share initial conditions, most of this sample variance cancels in their difference. The residuals therefore measure the model error relative to the precision of a $3\,h^{-3}\,\mathrm{Gpc}^3$ survey rather than a goodness of fit. 

In real space, statistical isotropy implies that the quadrupole and hexadecapole vanish in the ensemble average. In the mock, they are non-zero because finite-volume fluctuations and tracer discreteness. The model and the mock share the former but not the latter. Therefore, fitting all three multipoles tests whether the model reproduces these realisation-specific anisotropy up to discreteness noise. For $r > 35\,h^{-1}\mathrm{Mpc}$, the model follows both PRE and POST, including the sharpening of the BAO feature after reconstruction. The residuals rarely exceed $1\sigma$, for both the independent and the JOINT fits.

In redshift space, peculiar velocities generate a genuine quadrupole and hexadecapole. Reconstruction changes their shape while sharpening the BAO feature in the monopole. The reconstructed hybrid model reproduces all three multipoles over the full fitted range, with residuals below $1\sigma$. For this sample, the operator-based construction therefore remains accurate when reconstruction is combined with the redshift-space mapping.

\subsection{Scale cuts and choice of data vector}
\label{sec:scalecuts}

\begin{figure*}[!t]
    \centering
    \includegraphics[width=0.8\textwidth]{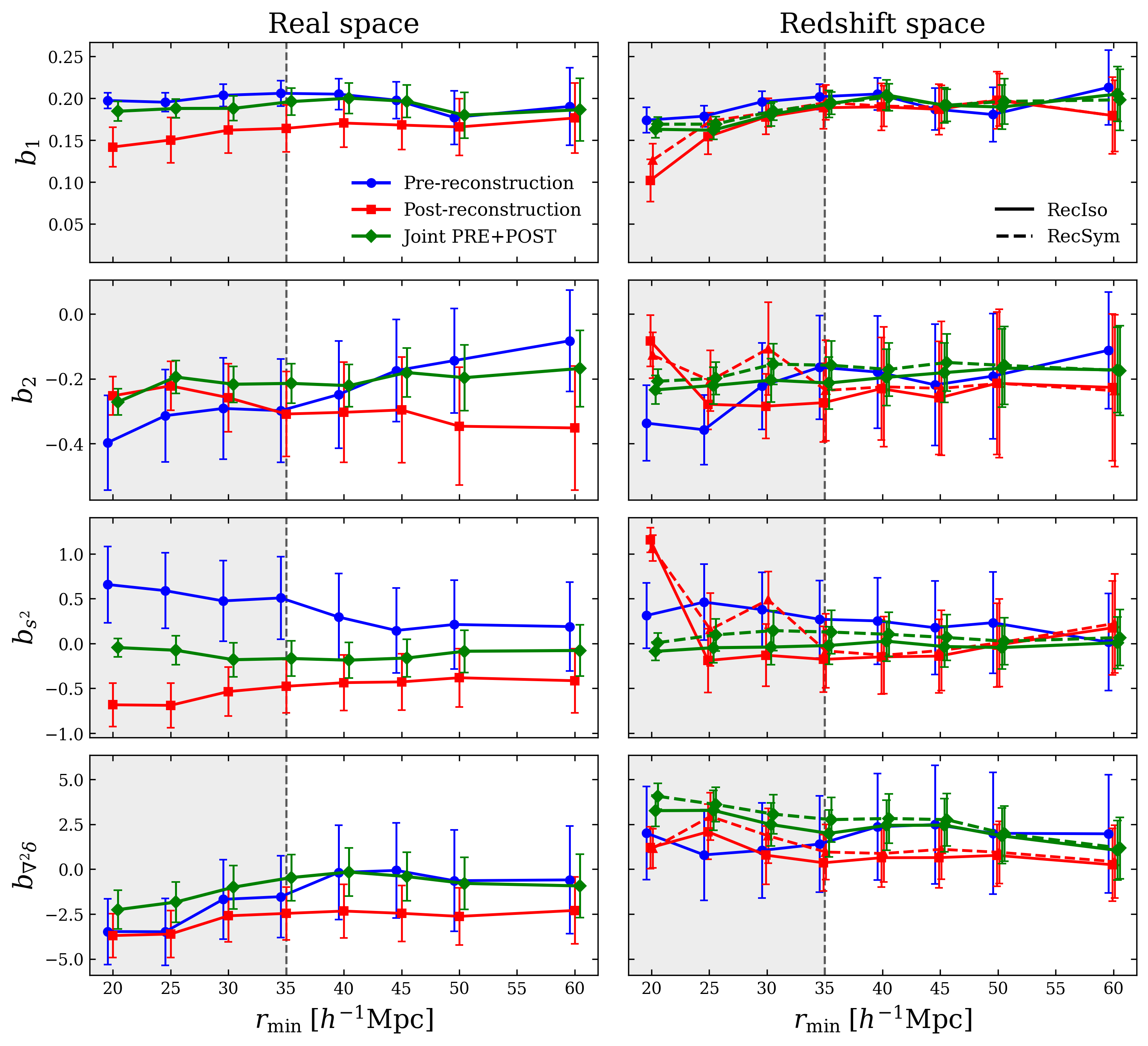}
    \caption{Best-fitting Lagrangian bias parameters as a function of the minimum separation $r_{\rm min}$, for PRE (blue), POST (red), and JOINT (green) fits to $\{\xi_0,\xi_2,\xi_4\}$. The left column shows the real-space results before and after reconstruction, while the right column shows the redshift-space results, with solid lines for RecIso and dashed lines for RecSym. Error bars show the $1\sigma$ parameter uncertainties from the corresponding fits, computed using the ensemble covariance. The vertical dashed line marks our fiducial $r_{\min}=35\,\hMpc$, and the grey region indicates the excluded range.}
    \label{fig:rmin_stability}
\end{figure*}

Figure~\ref{fig:rmin_stability} shows the best-fitting Lagrangian bias parameters as the minimum separation $r_{\min}$ varies from $20$ to $60\,\hMpc$. Below $30\,\hMpc$, the parameters drift with $r_{\min}$, most strongly for the POST redshift-space fits. Above it, they are stable within their uncertainties. For $r_{\min}\geq35\,\hMpc$, the largest shift relative to the $r_{\min}=60\,\hMpc$ fit is $\sim1.4\sigma$ for the individual fits and $\sim0.9\sigma$ for the JOINT fits, in units of the $r_{\min}=60\,\hMpc$ uncertainties. Therefore, we adopt $r_{\rm min} = 35\,h^{-1}\mathrm{Mpc}$ as our fiducial scale cut.

The cut is more conservative than those found in previous validations of the BACCO hybrid model in redshift space \citep{Zennaro2023, pellejero2023, ibanez2026validation}. The comparison is, however, only qualitative since these analyses were performed in Fourier space and modelled the virial motions of satellites, which we omit. The stronger drift in redshift space suggests that part of it comes from this omission, and we expect a small-scale Finger-of-God term to allow smaller $r_{\rm min}$. 

\subsection{Single bias set for PRE and POST}
\label{sec:singlebias}

A key test of our hybrid model is whether PRE and POST can be described with the same bias parameters. In real space and for each redshift-space convention, we compare two fits to PRE+POST data vector with the full covariance: a JOINT fit with a single set of bias parameters, and an independent fit with separate bias sets for PRE and POST. 

Figure~\ref{fig:multipoles} shows that the JOINT and independent fits are nearly indistinguishable in both real and redshift space. The JOINT fits give $\chi^2=45.9$, $61.8$, and $58.9$ in real space, RecIso, and RecSym, respectively, for 146 nominal degrees of freedom. These values are low because the mock and the model share initial conditions and the fixed-and-paired average suppresses sample variance, while the covariance describes single independent volumes. Allowing independent PRE and POST bias parameters reduces these values by only $\Delta\chi^2=1.59$, $1.61$, and $0.83$. In real space, the independent bias best-fit values differ but they sit along the same degeneracy directions, thus this does not translate into a substantial improvement in the fit. 

The stability of the fitted parameters with $r_{\rm min}$ (Fig.~\ref{fig:rmin_stability}) further supports the consistency of the independent and JOINT descriptions over the fitted range. Therefore, within the precision and scale range considered here, the data does not require requires a separate Lagrangian bias expansion after reconstruction.

\subsection{Galaxy-driven reconstruction}
\label{subsec:robustness}

In a real analysis, the displacement is estimated from the observed tracers, not from the matter field. That estimate carries the nonlinear bias and shot noise of the tracer and depends on the assumed $b_{\rm rec}$, whereas our model basis uses the matter displacement to keep the factorisation of Eq.~(\ref{eq:xi_rec_model}). To test this simplification, we reconstruct the \texttt{SHAMe} mock from its own galaxy field, with $b_{\rm rec} = 1.2$ and otherwise fiducial settings, and compare it with our matter-reconstructed model. Keeping the bias parameters fixed at their  JOINT best fit values, $\chi^2$ increases by only $2.68$ for RecIso and $0.42$ for RecSym over the full JOINT data vector. Since the PRE measurements are unaffected, the whole change comes from POST. 

The smaller change for RecSym is expected. With $d_S=d_D$, the displacement cancels in $\delta_{\rm rec}$ at linear order. In contrast, for RecIso, the line-of-sight component of the shift is not cancelled, and displacement errors enter already at linear order. At the precision of our tests, a basis reconstructed with the matter displacement describes data reconstructed from the galaxies themselves. RecSym is the more robust convention for this purpose.

\section{Cosmological constraints}
\label{sec:fisher}

We now compare the constraints on $\sigma_8$ and $\Omega_{\rm m}$ from PRE, POST, and JOINT using the Fisher-matrix formalism \citep{tegmark1997}. Our setting corresponds to a sample with $\bar{n}=5\times10^{-4}\,h^3\,{\rm Mpc}^{-3}$
in a simulation volume $V=(1440\,h^{-1}{\rm Mpc})^3 \simeq 3\,h^{-3}\,{\rm Gpc}^3$ at $z\simeq0$, with all other cosmological parameters fixed and no Alcock–Paczynski distortions. These results should therefore be regarded as a comparison between statistics under idealised conditions. Unless otherwise stated, we use the redshift-space $\{\xi_0,\xi_2,\xi_4\}$ data vector with $r_{\min}=35\,\hMpc$.

\subsection{Fisher formalism}
\label{subsec:fisher_setup}

\begin{figure*}[!t]
    \centering
    \includegraphics[width=0.9\textwidth]
    {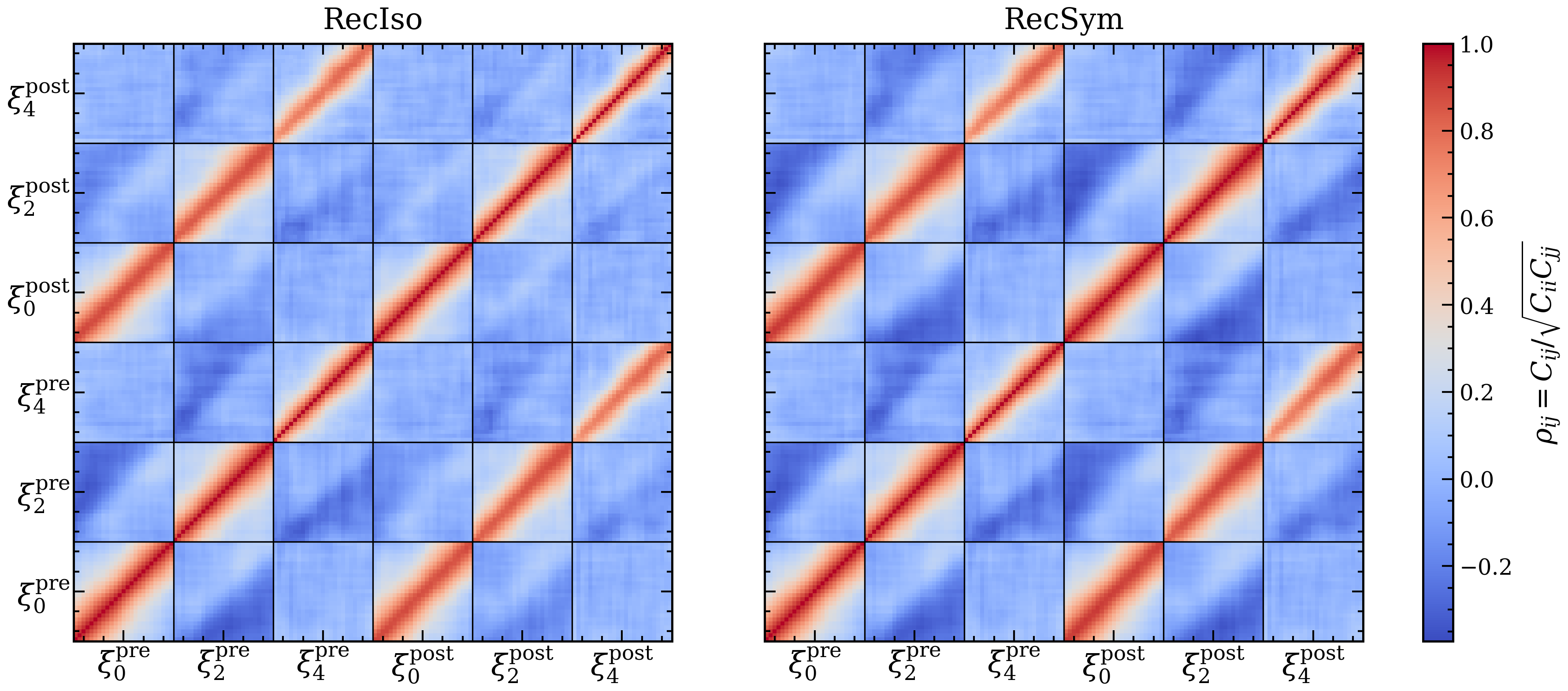}
    \caption{Correlation matrices of the joint $\{\xi_0,\xi_2,\xi_4\}_{\rm PRE+POST}$ data vector for RecIso (left) and RecSym (right) at $r>35\,h^{-1}{\rm Mpc}$, estimated from the covariance ensemble. The first three blocks correspond to the PRE multipoles, and the last three to their POST counterparts. The off-diagonal PRE--POST blocks encode the cross-covariance included in the JOINT Fisher analysis.}
    \label{fig:joint_correlation_matrix}
\end{figure*}

\begin{figure*}[!t]
    \centering
    \includegraphics[width=0.9\textwidth]{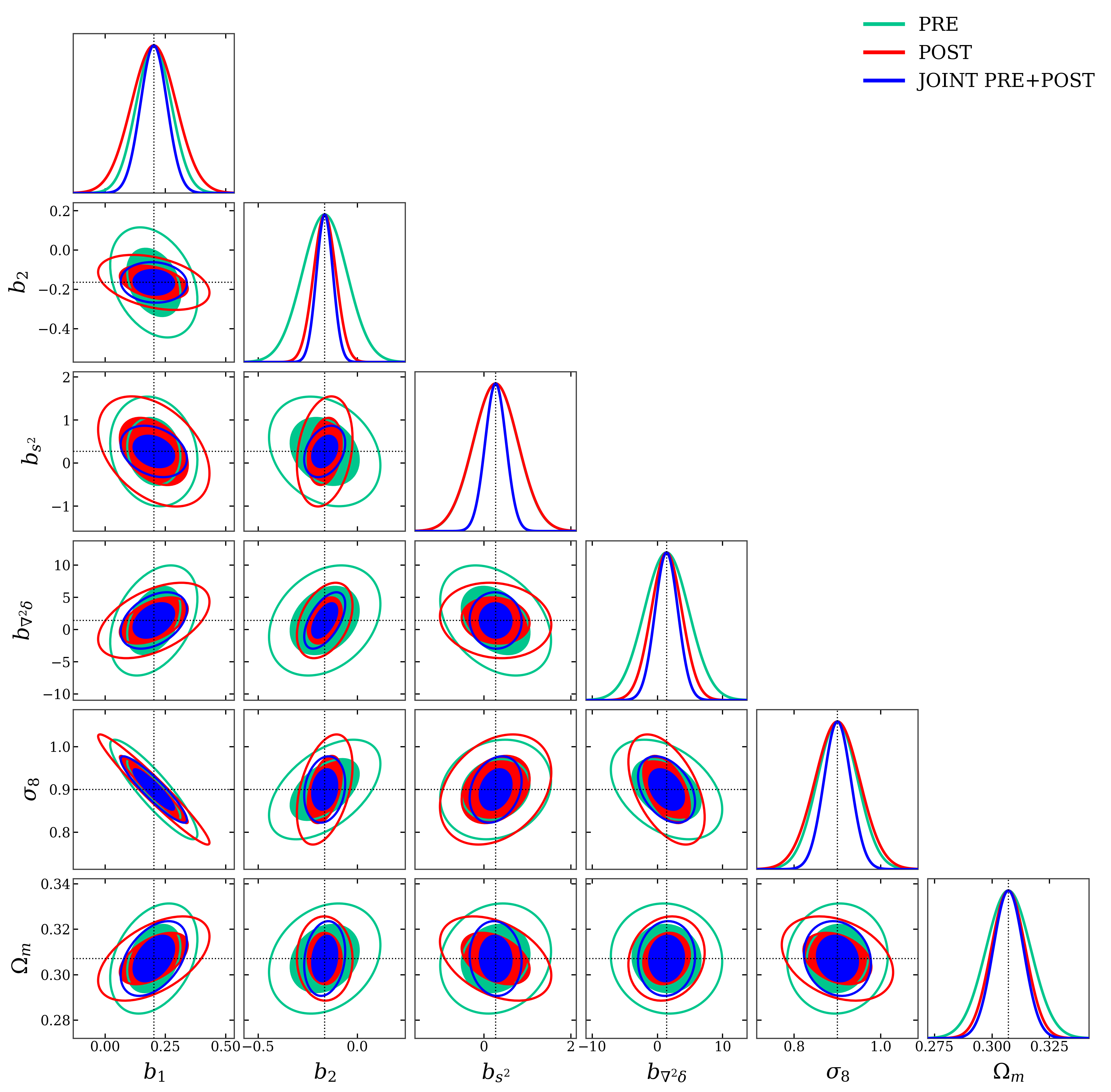}
    \caption{Marginalised Fisher constraints on $\{b_1,b_2,b_{s^2},b_{\nabla^2\delta},\sigma_8,\Omega_{\rm m}\}$ for the fiducial $\{\xi_0,\xi_2,\xi_4\}$ data vector using the RecSym reconstruction convention. Green, red, and blue contours correspond to PRE, POST, and the joint PRE+POST analysis, respectively. The filled regions show the 68\% confidence regions, while the contours enclose the 68\% and 95\% regions. All forecasts use the same fiducial point (dotted lines), given by the best-fitting PRE bias parameters. The POST and JOINT derivatives include the parameter dependence of the reconstruction displacement.}
    \label{fig:fisher_corner}
\end{figure*}

The JOINT data vector concatenates PRE and POST, $\mathbf{d}_{\rm joint} =(\mathbf{d}_{\rm pre},\mathbf{d}_{\rm post})^{\rm T}$. Since both are measured from the same realisations, their covariance has the block structure
\begin{equation}
    \mathbf{C}_{\rm joint}
    =
    \begin{pmatrix}
        \mathbf{C}_{\rm pre} & \mathbf{C_X}\\
        \mathbf{C_X}^{\rm T}   & \mathbf{C}_{\rm post}
    \end{pmatrix},
    \qquad
    \mathbf{C_X}
    \equiv
    {\rm Cov}
    \left(
        \mathbf{d}_{\rm pre},
        \mathbf{d}_{\rm post}
    \right),
    \label{eq:joint_covariance}
\end{equation}

\noindent where all blocks are estimated from the covariance ensemble
(Sect.~\ref{subsec:emulated}). 

Figure~\ref{fig:joint_correlation_matrix} shows the corresponding correlation matrices. The strongest correlations are between the same multipole at neighbouring separations, but appreciable correlations also extend across multipoles and between PRE and POST. Therefore, we include the full cross-covariance throughout.

Our JOINT data vector contains only the PRE and POST multipoles. Their statistical dependence is fully accounted for through the cross-covariance block $\mathbf{C_X}$, but we do not include the PRE--POST cross-correlation as a third observable. Unlike analyses that explicitly include the cross-power spectrum \citep{wang2024extracting, wang2024emulating, zhang2025cosmological}, we focus on isolating the complementarity between PRE and POST alone. 

Assuming a Gaussian likelihood with a parameter-independent covariance, the Fisher matrix for $\boldsymbol{\theta}= \{b_1,b_2,b_{s^2},b_{\nabla^2\delta},\sigma_8,\Omega_{\rm m}\}$ is

\begin{equation}
\label{eq:jb}
    \mathbf{F}
    =
    \mathbf{J}
    \mathbf{C}_{\rm H}^{-1}
    \mathbf{J}^{\rm T},
    \qquad
    J_{\alpha m}
    \equiv
    \frac{\partial d_m}{\partial\theta_\alpha},
\end{equation}

\noindent where $\mathbf{C}_{\rm H}^{-1}$ is the Hartlap-corrected precision matrix.

All Fisher matrices are evaluated at a single fiducial point: the cosmology of the simulation, $(\sigma_8,\Omega_{\rm m})=(0.9,\,0.307)$, and the bias parameters of the PRE fit, $(b_1,b_2,b_{s^2},b_{\nabla^2\delta}) = (0.201,\,-0.164,\,0.271,\,1.401)$. The reconstruction growth rate $f_{\rm rec}$ is held fixed throughout the cosmological derivatives.

We define the figure of merit (FoM) as:

\begin{equation}
    {\rm FoM}
    \equiv
    \left[
        \det \mathbf{C}^{(\sigma_8,\Omega_{\rm m})}
    \right]^{-1/2},
    \label{eq:fisher_fom}
\end{equation}

\noindent where $\mathbf{C}^{(\sigma_8,\Omega_{\rm m})}$ is the marginalised $2\times2$ sub-block of $\mathbf{F}^{-1}$. To isolate the impact of bias marginalisation, we also quote constraints with the bias parameters held fixed. These are obtained from the inverse of the cosmological sub-block of $\mathbf{F}$.

\subsection{Parameter dependence of the reconstruction}

In a real analysis, reconstruction is applied to the observed tracer field with a fiducial cosmology, a fixed $f_{\rm rec}$, and a value of $b_{\rm rec}$ inferred from the observed clustering amplitude. The Fisher derivatives must therefore describe how the expected data vector responds to the true parameters. Two ingredients enter: the dependence of the tracer field being on the true parameters, and the way $b_{\rm rec}$ is chosen.

Consider a sample with linear bias $b^E_1$, so that $\delta_g \propto b^E_1 \sigma_8$. In practice, $b_{\rm rec}$ is derived from the observed clustering amplitude assuming a fiducial $\sigma_8$,

\begin{equation}
    b_{\rm rec} = \frac{\left(b^E_1\,\sigma_8\right)_{\rm obs}}{\sigma_{8,\rm fid}}.
    \label{eq:cosmo_rec}
\end{equation}

At linear order, the real-space reconstruction input is then $\delta_g/b_{\rm rec} = \delta_m\,\sigma_{8,\rm fid}/\sigma_8$. That is, the linear bias cancels, and the estimated displacement has the amplitude it would have if the true $\sigma_8$ were $\sigma_{8,\rm fid}$, independently of the true value of $b_1$.

Our hybrid reconstruction model uses the matter field with $b_{\rm rec} = 1$. If the fiducial and true cosmology coincide, this is equivalent at linear order to reconstructing the tracer field with $b_{\rm rec} = b^E_1$. When $\sigma_{8,\rm fid}$ differs from the true $\sigma_8$, Eq.~(\ref{eq:cosmo_rec}) is instead equivalent to reconstructing the matter field with

\begin{equation}
b_{\rm rec} = \frac{\sigma_8}{\sigma_{8,\rm fid}}. 
    \label{eq:cosmo_rec_s8}
\end{equation}

In redshift space, the equivalence at the fiducial point requires $f_{\rm rec}$ to be equal to the true growth rate, and the response to $\sigma_8$ is reproduced only approximately. For our sample, this overestimates the line-of-sight response by $\sim 10\%$ at $\mu=1$. Thus, we expect our treatment to capture the leading-order parameter dependence of the reconstruction. 

We compute all bias derivatives analytically from the reconstructed basis correlations of the emulated fiducial realisations. For the $\Omega_{\rm m}$ derivatives, we reconstruct the matter field at each endpoint with $b_{\rm rec} = 1$, so that the displacement responds to $\Omega_{\rm m}$ only through the shape of the matter field. For the $\sigma_8$ derivatives, we use $b_{\rm rec}$ given by Eq.~(\ref{eq:cosmo_rec_s8}).

We evaluate the $\sigma_8$ and $\Omega_{\rm m}$ derivatives using centred finite differences with $\Delta\sigma_8=0.045$ and $\Delta\Omega_{\rm m}=0.015$. Our forecasts use Richardson-extrapolated derivatives, $J=(4J_{\Delta/2}-J_{\Delta})/3$, built from the full and half steps, and we have checked that our results do not change if we instead use the half-step derivatives alone.

\subsection{Pre- versus post-reconstruction}

\begin{table}
\caption{Fisher gains relative to PRE for the fiducial $\{\xi_0,\xi_2,\xi_4\}$ data vector. We define $G_{\sigma_8}\equiv \sigma_{\rm PRE}(\sigma_8)/\sigma(\sigma_8)$, $G_{\Omega_{\rm m}}\equiv \sigma_{\rm PRE}(\Omega_{\rm m})/\sigma(\Omega_{\rm m})$, and $G_{\rm FoM}\equiv{\rm FoM}/{\rm FoM}_{\rm PRE}$, so that values larger than unity indicate an improvement over PRE. Values in parentheses are gains with the four Lagrangian bias parameters held fixed, normalised to the PRE case. All results are evaluated at the same fiducial point and use the Hartlap-corrected precision matrices. The marginalised PRE reference values are $\sigma(\sigma_8)=0.04681$, $\sigma(\Omega_{\rm m})=0.00974$, and ${\rm FoM}=2195.63$. For fixed biases, the corresponding PRE values are $0.01396$, $0.00687$, and $16447.57$.}
\label{tab:fisher_constraints}
\centering
\begin{tabular}{@{}llccc@{}}
\hline\hline
Conv. & Stat.
& $G_{\sigma_8}$
& $G_{\Omega_{\rm m}}$
& $G_{\rm FoM}$ \\
\hline
--     & PRE   & 1.00 (1.00) & 1.00 (1.00) & 1.00 (1.00) \\
RecIso & POST  & 0.97 (1.24) & 1.27 (1.29) & 1.29 (1.40) \\
       & JOINT & 1.51 (1.25) & 1.50 (1.30) & 2.27 (1.44) \\
RecSym & POST  & 0.90 (1.28) & 1.31 (1.31) & 1.25 (1.48) \\
       & JOINT & 1.50 (1.28) & 1.47 (1.30) & 2.23 (1.50) \\
\hline
\end{tabular}
\end{table}

Figure~\ref{fig:fisher_corner} shows the full six-dimensional Fisher constraints for the RecSym convention and Table~\ref{tab:fisher_constraints} lists the corresponding gains relative to PRE. 

For RecSym, POST improves the marginalised constraint on $\Omega_{\rm m}$ by a factor of $1.31$ relative to PRE, plausibly due to the sharper BAO feature after reconstruction. The marginalised constraint on $\sigma_8$, in contrast, is about 10\% weaker ($G_{\sigma_8} = 0.90$), but the $(\sigma_8, \Omega_{\rm m})$ FoM still increases by a factor of $1.25$. RecIso behaves similarly, with $G_{\Omega_m} = 1.27$, $G_{\sigma_8} = 0.97$ and a FoM gain of 1.29.

With the bias parameters fixed, POST constrains $\sigma_8$ 1.28 times better than PRE (1.25 for RecIso). This indicates that POST carries more direct information on $\sigma_8$. Its weaker marginalised constraint thus reflects a stronger degeneracy between $\sigma_8$ and the bias parameters. 

\subsection{Combining PRE and POST}
\label{sec:joint}

Combining PRE and POST improves the marginalised cosmological constraints for both reconstruction conventions (Table~\ref{tab:fisher_constraints}). For RecSym, JOINT tightens the constraints on $\sigma_8$ and $\Omega_{\rm m}$ by a factor of 1.50 relative to PRE, which reduces each uncertainty by about one third, and increases the FoM by a factor of 2.23. RecIso gives gains within 2\% of these. Hence, under our reconstruction prescription, the choice of convention has little impact on the combined constraints.

At fixed bias, JOINT increases the FoM by a factor of only 1.50 (RecSym) and 1.44 (RecIso), almost the same as POST alone (1.48 and 1.40), so PRE adds little to POST. After marginalisation, however, the JOINT gain rises to 2.23--2.27, compared with 1.25--1.29 for POST. A large part of the JOINT improvement therefore comes from breaking degeneracies between bias and cosmology. Figure~\ref{fig:fisher_corner} shows the bias-$\sigma_8$ degeneracy has different orientations in PRE and POST, so their combination constraints the bias better than either statistics alone, which in turn tightens the cosmological constraints.  

\subsection{Comparison with previous work}

Our results agree qualitatively with previous studies showing that pre- and post-reconstruction clustering contain complementary information \citep{wang2024extracting, wang2024emulating, zhang2025cosmological}. 

\citet{wang2024emulating} found that jointly analysing the pre-, post-reconstruction, and cross-power spectra reduced the uncertainties in $\Omega_{\rm m}$, $H_0$, and $\sigma_8$ by approximately $41$--$55\%$ relative to the pre-reconstruction power spectrum alone, using $k_{\rm max}=0.25\,h\,{\rm Mpc}^{-1}$. More recently, \citet{zhang2025cosmological} reported improvements of about $11$--$16\%$ in parameter constraints relative to POST alone, together with a $10\%$ increase in the FoM, when combining the pre-, post-, and cross-power spectra up to $k_{\rm max}=0.2\,h\,{\rm Mpc}^{-1}$. Applied to DESI DR1 LRGs, \citet{wang2026nonlinear} found $18$--$27\%$ improvements in the $\sigma_8$ constraints relative to a pre-reconstruction-only analysis with the same external priors.

These improvements are comparable to our reduction in the marginalised uncertainty on $\sigma_8$ for JOINT relative to PRE. A direct comparison is nevertheless not possible: the analyses differ in their observables, parameters, scales and modelling, and our idealised setting fixes all other cosmological parameters and omits Alcock–Paczynski distortions, which tends to enlarge the gains. 

Notably, our gains are obtained without including the PRE-POST cross-correlation as an additional observable. Because PRE and POST share a single hybrid bias expansion, the cross-correlation could be added without new bias parameters, and the field-level formulation accommodates different reconstruction conventions directly.

\section{Conclusions}
\label{sec:conclusions}

We have extended the hybrid Lagrangian bias framework to post-reconstruction galaxy clustering. The central idea is to apply a reconstruction displacement to each Eulerian basis field of the hybrid expansion, in the same way as to the galaxy catalogue. The reconstructed galaxy correlation function is then a combination of 15 reconstructed basis correlations, weighted by the same Lagrangian bias parameters that describe pre-reconstruction clustering. This separation between cosmology-dependent functions and tracer-dependent bias coefficients makes the framework particularly suitable for emulation, without adopting a specific model of the galaxy--halo connection. 

We validated our framework against \texttt{SHAMe} mocks built from the BACCO simulations at $z \simeq 0$, for a sample with $\bar n = 5\times10^{-4}\,h^{3}\mathrm{Mpc}^{-3}$. In real and redshift space, the model reproduces the monopole, quadrupole, and hexadecapole before and after reconstruction for $r>35\,h^{-1}{\rm Mpc}$, with a single set of Lagrangian bias parameters. This shows that, for the scales and galaxy population studied here, reconstruction acts as a transformation of the hybrid basis fields and does not require a different bias expansion. Our scale cut is set partly by the omission of satellite virial velocities; including them should allow the model to reach smaller scales.

Using a Fisher formalism, POST alone tightens the marginalised $\Omega_{\rm m}$ constraint by a factor of $\simeq1.3$ relative to PRE, but slightly weakens the marginalised $\sigma_8$ constraints due to a stronger bias--amplitude degeneracy. Combining PRE and POST reduces both marginalised constraints by about one third and increases the FoM by a factor $\sim2.2$. RecIso and RecSym give  comparable JOINT constraints, demonstrating that this complementarity does not depend on the convention under our prescription.

Applying this model to a real survey requires additional aspects. First, reconstruction assumes a fiducial cosmology to convert observed angles and redshifts into comoving coordinates, which introduces Alcock--Paczynski distortions \citep{AlcockPaczynski1979}. Since our predictions are three-dimensional fields, these can be forward modelled, as in emulator-based analyses \citep[e.g.][]{wang2024emulating,wang2026nonlinear}. Second, the displacement must be estimated from a noisy tracer field, assuming fiducial values for the growth rate and the linear bias, which breaks the exact factorisation between basis, bias parameters, and stochastic contributions. Our galaxy-driven test shows that the impact of these is small, but they should nevertheless be assessed for the target sample and required accuracy. Third, the forecast gains should be re-evaluated with the full set of cosmological parameters free.

Future work should emulate the reconstructed basis as a function of cosmology, add a small-scale velocity term to extend the fitted range, include the PRE--POST cross-correlation as an additional observable, test other tracer populations and reconstruction settings, and incorporate observational effects. Extending the analysis to full-likelihood inference will test whether the forecast PRE--POST complementarity holds beyond the Fisher approximation.

\begin{acknowledgements}
VID acknowledges support from the Dirección de Postgrado of Universidad Técnica Federico Santa María through the Master's scholarship programme. REA received support from grant PID2024-161003NB-I00 funded by MICIU/AEI/10.13039/501100011033 and by ERDF/EU. ADMD and VID acknowledge support from the Universidad Técnica Federico Santa María through the Proyecto Interno Regular \texttt{PI\_LIR\_25\_04}.
\end{acknowledgements}

\bibliographystyle{aa}
\bibliography{references}

\end{document}